# Non-Equilibruim Molecular Dynamics Simulation of Poiseuille Flow in a Carbon Nanochannel

Ni Guo Liang, He Ming Li, Hua Yao Zu, Bagher Abareshi

## Abstract

*The numerical simulation of a Poiseuille flow in a narrow channel using the molecular dynamics simulation (MDS) is performed. Poiseuille flow of liquid Argon in a carbon nanochannel is simulated by embedding the fluid particles in a uniform force field. Density, velocity and Temperature profiles across the channel are investigated. When particles will be inserted into the flow, it is expected that the dynamics of flow will depend on the thermostat chosen. To obtain a more uniform temperature distribution across the channel we use local thermostating near the wall. The obtained results show that velocity profile, slip length and slip velocity depends on the driving force.*

*Keywords:* Molecular Dynamics Simulation, Poiseuille Flow, Nanochannel, Velocity Profile, Slip length, Slip Velocity.

## 1. Introduction

Computer simulation is a tool for studying macroscopic systems by implementing microscopic models. Molecular dynamics merges computer simulation with statistical mechanics to compute equilibrium and transport properties of a classical many-body system. The main idea behind the classical MDS is the calculation of the interactions between atoms and the solution of Newton's equations of motion for each particle in order to extract thermothynamics properties [1]. One must bear in mind that though molecular dynamics aims to study microscopic systems, only a finite number of particles, from a few hundred to a few million, can be simulated on a computer.

For atomic system far from equilibrium, such as liquid flows, non-equilibrium molecular dynamics (NEMD) offers an effective simulation method as well as an alternative method for the calculation of transport properties of liquids[2]. During recent years, NEMD techniques have dominated the field of nano- fluidics simulation [3]. Researchers have dealt with the question whether the classical Navier-Stokes equations are valid in channels of small widths. The observations show that densities profils reveal strong oscillations in the number of fluid atoms at layers adjacent to the walls.

Hansen and Ottesen [4] noted that Navier- stokes hydrodynamic prediction for the velocity profile break down at small channel width below 5 molecular diameters. Somers and Davis [5] presented an early full analysis of density and velocity profiles for channel width from 0.6 nm to 0.8 nm. Another critical aspect in nano-and micro-flows is the existence of slip at the solid surfaces. Galea and Attard [6] Concluded that the no slip condition breaks down in small channel widths. Temperature profiles in NEMD simulations of channel flow show a deviation from the continuum-theory predictions, i.e uniform profile when a viscous dissipation is negligible and quadratic profile when viscous dissipation is accounted for. Akhmatskaya et al. [7] extracted the temperature profiles for channel width of 1.5nm. Near the walls, they reported a rise in temperature and in the middle of the channel fluid temperature is constant, although it is higher than the wall temperature. Delhommele and Evans [8] extracted the temperature profile for a $4\sigma$ ideal atomic system and found a flat profile for two different external forces, while they obtained flat temperature profiles in simulation of real fluids in a $10.2\sigma$ channel. Sofos et al. [9] presented NEMD of planner poiseuiulle flow of liquid Argon. They showed that the contribution of parameters such as temperature and external forces are not negligible at nanoscale. At the atomic scale, the detailed investigation and categorization of all possible parameters that affect nano-flow phenomena is of a particular interest. In this scale, the effect of wall on fluid atom is strong and the continuum theory breaks down.

The purpose of this paper is to examine the effect of external forces on the velocity profiles in order to better understanding the fluidic flow behavior in the nano channel geometry. We have carried out NEMD simulations of simple Lennard-jones fluids subject to an inlet driving force in a nano-channel. In the present work, to obtain a more uniform temperature distribution across the channel, we use local thermostatting near the wall. In this method, the wall atoms are assumed to have infinite mass and therefore remain static at their original positions throughout the simulation. This does not allow for heat exchange or control via the walls. In this case, thermostats that artificially extract heat from the system e.g. velocity rescaling is used, together with the velocity integrator for integration of the trajectories of the particles. For this purpose, at the near wall region, along the flow- direction on the both sides, the thermostat acts in each of them. The interfacial density profiles at different magnitude of the driving force are obtained. Also the effect of external force on the velocity profile and slip velocity are examined.

## 2. Simulation methods

In this paper, molecular dynamics simulations are performed for liquid molecules moving in a nano-channel bounded by two solid walls on the x-axis, Subject to a driving force at the inlet as shown in Fig.1. The simulation cell has the size of $17\sigma \times 12\sigma \times 21\sigma$, where $\sigma$ is a size parameter that represents the molecule's diameter. For the sake of physical understanding, we assume that the imaginary solid walls are made of platinum and Lennard-Jones (LJ) fluid is argon. The solid walls are located at $z=0$ and $z=L_z$, where $L_z$ is the box length along z direction. For this geometry, there are 2700, 2550 and 2400 Argon molecules (at its saturated liquid density) at 102 K, 108 K and 120 K respectively, inside the carbon nanochannel. Periodic boundary conditions are applied along the x-and y-directions.

The atomic interaction is governed by the modified Lennard-Jones potential as defined by Stoddard and Ford [10]:

$$\phi(r_{ij}) = \left\{ 4\varepsilon \left[ \left(\frac{\sigma}{r_{ij}}\right)^{12} - \left(\frac{\sigma}{r_{ij}}\right)^{6} \right] \right\} + \left\{ 4\varepsilon \left[ 6\left(\frac{\sigma}{r_c}\right)^{12} - 3\left(\frac{\sigma}{r_c}\right)^{6} \right] \left(\frac{r_{ij}}{r_c}\right)^{2} \right\} - \left\{ 4\varepsilon \left[ 7\left(\frac{\sigma}{r_c}\right)^{12} - 47\left(\frac{\sigma}{r_c}\right)^{6} \right] \right\} \tag{1}$$

Where $r_{ij}$ is the inter-particle separation, $i$ and $j$ denote the particles $i$ and $j$, respectively, $\varepsilon$ is the maximum potential depth and it is a parameter corresponding to the energy, $\sigma$ is the size parameter and $r_c$ is the cut-off radius beyond which the intermolecular interaction could be ignored. Hence, for the purpose of physical understanding, in the present study, argon is used as the LJ fluid with the following potential parameters: $m = 6.63 \times 10^{-26} kg$, $\sigma = 3.045 A°$ and $\varepsilon = 1.67 \times 10^{-21} J$. The solid wall is represented by one layer of face centered cubic surface of platinum molecules with parameters as: $m_s = 3.24 \times 10^{-26} kg$, $\sigma_s = 2.475 A°$ and $\varepsilon_s = 8.35 \times 10^{-20} J$.

To study the wall effects on the surface tension and density profile we placed two solid walls, one at bottom boundary and the other at the up boundary. The potential function between the solid and liquid molecules is represented by Lennard-Jones function as [10]:

$$\phi_w(r_{ij}) = \left\{ 4\varepsilon_{sf} \left[ \left(\frac{\sigma_{sf}}{z}\right)^{12} - \left(\frac{\sigma_{sf}}{z}\right)^{6} \right] \right\} + \left\{ 4\varepsilon_{sf} \left[ 6\left(\frac{\sigma_{sf}}{r_c}\right)^{12} - 3\left(\frac{\sigma_{sf}}{r_c}\right)^{6} \right] \left(\frac{z}{r_c}\right)^{2} \right\} - \left\{ 4\varepsilon_{sf} \left[ 7\left(\frac{\sigma_{sf}}{r_c}\right)^{12} - 47\left(\frac{\sigma_{sf}}{r_c}\right)^{6} \right] \right\} \tag{2}$$

$$\sigma_{Sf} = \left[ \frac{1}{2^{13}} \frac{\varepsilon \sigma^{12}}{\sqrt{\varepsilon \sigma^6 \varepsilon_s \sigma_s^6}} \left( 1 + \left(\frac{\varepsilon_s \sigma_s^{12}}{\varepsilon \sigma^{12}}\right)^{\frac{1}{13}} \right)^{13} \right]^{\frac{1}{6}} \tag{3}$$

$$\varepsilon_{sf} = \frac{\sqrt{\varepsilon \sigma^6 \varepsilon_s \sigma_s^6}}{\sigma_{sf}^6} \tag{4}$$

Where, the Kong mixing rules are applied to the system. When a two-body potential model is chosen the interaction force between a pair of molecules can be derived from potential by the following relation:

$$\vec{F}_i = \sum_{i \neq j} \vec{F}_{ij} + \vec{F}_{ext} \tag{5}$$
$$\vec{F}_{ij} = -\vec{\nabla}(\phi + \phi_w)$$

Where $\vec{F}_{ext}$ is the inlet external force per molecule. The cut-off radius for the LJ potential in all simulation is $r_c = 4\sigma$. An external force $\vec{F}_{ext}$, ranging from 1to11PN (PN= Pico Newton), is applied along the flow direction to inlet fluid particles during the simulation. In order to extract heat from the system, velocity rescaling (Berandson thermostat) is used for the bins near the walls. For velocity rescaling the velocities are first updated from the force acting on the particles and rescaled at each time step according to:

$$V_{new} = \lambda V_{old} \qquad where \qquad \lambda = \left(\frac{T_{t\arg et}}{T_{global}}\right)^{\frac{1}{2}} \tag{6}$$

Where $T_{target}$ is the target temperature and $T_{global}$ is the global temperature of the fluid particles in the system. The temperature is defined here as a fluctuation kinetic energy per particle at each slab, per degree of freedom;

$$T_{inst} = \frac{\frac{1}{2}m(V_x^2 + V_y^2 + V_z^2)}{3\ N(Z)\ K_B} \tag{7}$$

Where $N(z)$ is the number of atoms at height $z$. Atoms are given appropriate initial velocities in order to reach the desired temperature. The system reaches equilibrium state after a run of 200000 time steps. Then, NEMD simulations are performed with duration of time steps. In all simulation cases, the selected time step was 0.5 picoseconds and more than 2,000,000 time steps were used in sampling data[12-16].

### 3. Results and Dicussion

In order to compute the density, streaming velocity and temperature profiles, the domain is divided into $N_{bin}$ bins along the z-direction, each one of volume $L_x \times L_y \times \Delta z$, Where $L_x$ and $L_y$ are the box lengths along x and y directions, respectively and $\Delta z$ is equal to $\left(L_z/N_{bin}\right)$. If $N_{zi}$ is the number of molecules in $i_{th}$ bin of step j, $N_{sample}$ is the number of sampling and $m$ is the mass of each molecule, then the density is given using the following equation:

$$\rho(z) = \frac{N_{zi}m}{N_{sample}L_x L_y \Delta z} \tag{8}$$

In all simulation $N_{bin}$ is equal to 250. Fig.2 shows the effect of external force on the density profiles based on the NEMD simulation. Simulation where carried out with 2400 particles at 120K with six different external force (per molecule) of 1,3,5,7,9 and 11 PN at the inlet. It is shown that the driving force has no significant influence on the density profile. This is in agreement with results reported by Nagayama and Cheng [11]. Therefore, we can conclude that the interfacial density profiles are independent of the magnitude of the driving force. Also, due to strong solid-liquid interactions, the liquid molecules are distributed orderly in the neighborhood of the solid interface. This interface structure (i.e. the solid-like structure formed by the absorption layers of the argon atoms) and its oscillatory characteristics of the density profiles are in agreement with majority of earlier works [5],[12-14]. Also, due to the nanoscale character of the distance between the walls, the influence of wall adsorption on the density of the liquid is very noticeable.

Fig.3 shows the effect of temperature on streaming velocity profile at external force of 3 PN. To extract the velocity profiles, mean velocity is computed at each bin for each time step and all values are averaged. It is clear from the obtained results that an increase in temperature results is an increase of fluid mobility and as a result, in maximum and average velocity will rise.

Fig.4 shows the velocity profiles of the flow in the z-direction along the carbon nanochannel for different external driving forces. Simulation where carried out with 2400 particles at 120.9K with six different external force of 1,3,5,7,9 and 11 PN at the inlet. As mentioned, an increase in $F_{ext}$ can increase maximum and average velocity values. In order to compare velocity behavior as a function of $F_{ext}$, we present the velocity profile of Fig.4 for the same $r_c$, T, $\varepsilon_{sf}, \varepsilon, \sigma_{sf}$. As seen form this figure the velocity profiles have a similar shape under different driving farces and the mean and maximum velocity will increase with increasing driving force.

As seen from this figure, parabolic fits are applied to velocity profiles and they are used in the extraction of slip length and slip velocity. The slip length $L_s$ is calculated according to equation (4):

$$L_s = \frac{\Delta V}{\dot{\gamma}} \quad (9)$$

Where $\dot{\gamma}$ is the local shear rate in the fluid and $\Delta V$ is the measured difference in velocity between the wall and neighboring fluid. As shown in Fig.4 we can obtain the slip length by extrapolating the velocity profiles from a position in the fluid to where the velocity would vanish within the solid. By using of a linear Navier boundary condition [13], the solution for the slip velocity $V_s$ near the wall normalized by the centerline velocity $U_c$ is,

$$\frac{V_s}{U_c} = (1 + \frac{D_z}{4L_s})^{-1} \quad (10)$$

Where $D_z$ and $L_s$ are the distances shown in Fig.5. Table1 summarizes the result of slip length by fitting the velocity profiles. It is shown that, the slip length is a function of driving force and it is not constant. The variation of the slip length shows an upward trend with increasing inlet driving force. Fig.6 and table.2 show the variation of the slip velocity as a function of the inlet driving forces for different temperatures. According to the earlier discussion, the slip velocity is defined to be the velocity in the first bin near the solid wall. According to Fig.6, the slip velocity increases with increasing inlet driving force. This figure shows that no-slip boundary condition is not valid for various inlet driving forces, and only for an extremely large driving force, the no-slip condition is valid.

Temperature profiles across the channel are presented in figure 7. We see that the temperature is approximately parabolic with a temperature rise in the middle region compared to near the surfaces. This is a consequence of viscous heating in the fluid. The parabolic shape originates simply from the fact that the heat flux through a given plane is proportional to the distance from the channel center. Since the heat flux is proportional to the temperature gradient, a parabolic temperature profile is obtained upon integration. Moreover, the temperature profiles at higher external forces appear to shift up and become discontinuous across the wall fluid interface. This behavior indicates the presence of thermal slip in addition to the velocity slip described above.

## 4 Conclusions

The NEMD simulation of poiseuille flow of liquid argon in a carbon nanochannel is presented and the following conclusions may be drawn:
- The density profile is independent of the magnitude of the driving force.
- Velocity profiles are significantly affected by temperature and the magnitude of the external forces.

- Parabolic fits for the velocity profiles can be accomplished for a great range of external forces.
- The slip length and slip velocity are functions of the driving forces and they show an arising trend with increasing inlet driving force.
- The temperature profiles are parabolic which is in agreement with earlier knowledge about Poiseuille flow.

**Nomenclature**

$\vec{F}_{ext}$ : External force
$k$ : Boltzman Constant ($K^{-1}$)
$L_s$ : Slip Length (m)
$L_x$ : Box length along *x* direction (m)
$L_y$ : Box length along *y* direction (m)
$L_z$ : Box length along *z* direction (m)
$m$ : Mass of molecule (Kg)
$N(z)$ : Number density at height z (Non-Dimensional)
$N_{atm}$ : Number of atoms (Non-Dimensional)
$N_{bin}$ : Number of rectangular slabs (Non-Dimensional)
$N_{zi}$ : Number of atoms in a slab
$N_{sample}$ : Number of sampling (Non-Dimensional)
$r$ : Position vector (m)
$r_c$ : Cutoff Radius (m)
$r_{ij}$ : Inter-particle separation (m)
$T$ : Temperature (K)
$v$ : Velocity vector ($ms^{-1}$)
$z$ : Height (m)

***Greek symbols***

$\vec{\nabla}$ : Dell Gradient
$\Delta z$ : Thickness of rectangular slab (m)
$\sigma$ : Length parameter for Argon (m)
$\sigma_s$ : Length parameter for Platinum (m)
$\rho$ : Density ($kgm^{-3}$)
$\varepsilon$ : Energy parameter (J)
$\varepsilon_s$ : Energy parameter for platinum (J)
$\phi$ : Potential function (J)
$\phi_w$ : Wall potential function (J)

s : Solid
x : x-direction
y : y-direction
z : z-direction

## Tables

**Table1** The result of slip length.

| Temperature(K) | $F_{ext}=1PN$ | $F_{ext}=3PN$ | $F_{ext}=5PN$ | $F_{ext}=7PN$ | $F_{ext}=9PN$ | $F_{ext}=11PN$ |
|---|---|---|---|---|---|---|
| | $L_s (nm)$ | | | | | |
| Temperature=102.7K | 0.03 | 0.045 | 0.059 | 0.072 | 0.081 | 0.090 |
| Temperature=108.8K | 0.034 | 0.048 | 0.064 | 0.077 | 0.085 | 0.094 |
| Temperature=120.9K | 0.037 | 0.052 | 0.068 | 0.080 | 0.088 | 0.097 |

**Table 2** The results of velocity slip.

| Temperature(K) | $F_{ext}=1PN$ | $F_{ext}=3PN$ | $F_{ext}=5PN$ | $F_{ext}=7PN$ | $F_{ext}=9PN$ | $F_{ext}=11PN$ |
|---|---|---|---|---|---|---|
| | $V_s/U_c$ | | | | | |
| Temperature=102.7K | 0.0021 | 0.00326 | 0.0042 | 0.0052 | 0.0058 | 0.0060 |
| Temperature=108.8K | 0.0024 | 0.0034 | 0.0046 | 0.0055 | 0.0060 | 0.0068 |
| Temperature=120.9K | 0.00268 | 0.0037 | 0.0049 | 0.0058 | 0.0064 | 0.0070 |

## Figures

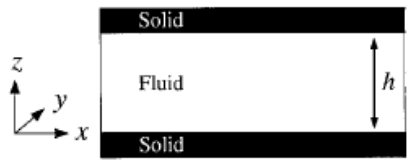

**Figure 1** Simulation system configuration of a nanochannel

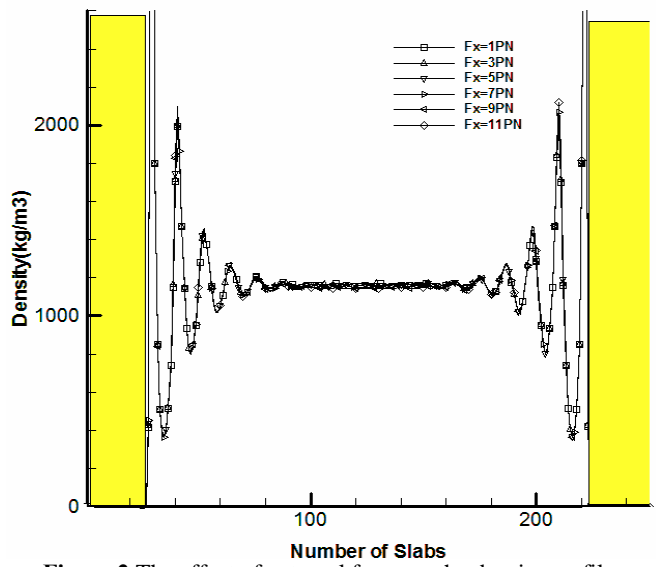
**Figure 2** The effect of external force on the density profiles

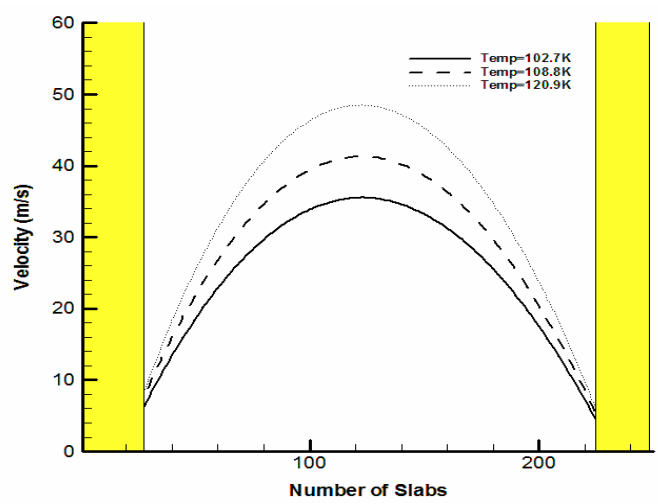
**Figure 3** The effect of temperature on streaming velocity profile

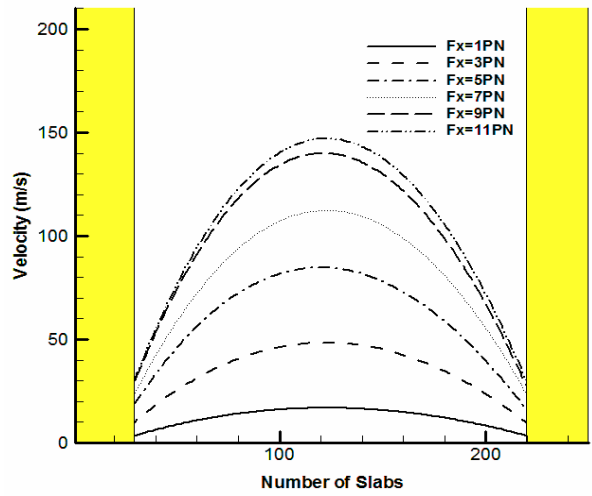

**Figure 4** The velocity profiles in the z-direction for different external driving forces

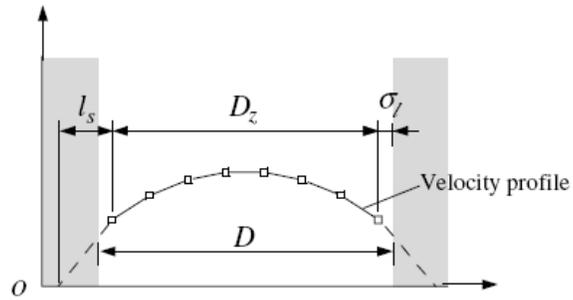

**Figure 5** Definition of the length of slip

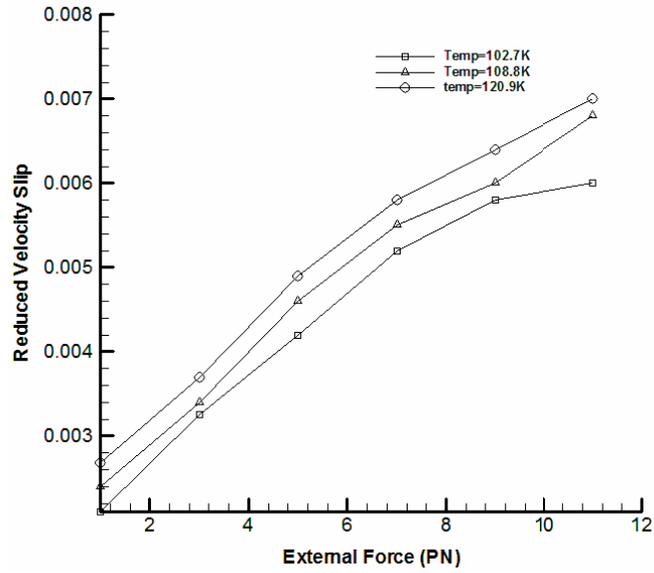

**Figure 6** Variation of the reduced slip velocity ($V_s/U_c$) as a function of the inlet driving force for different temperatures

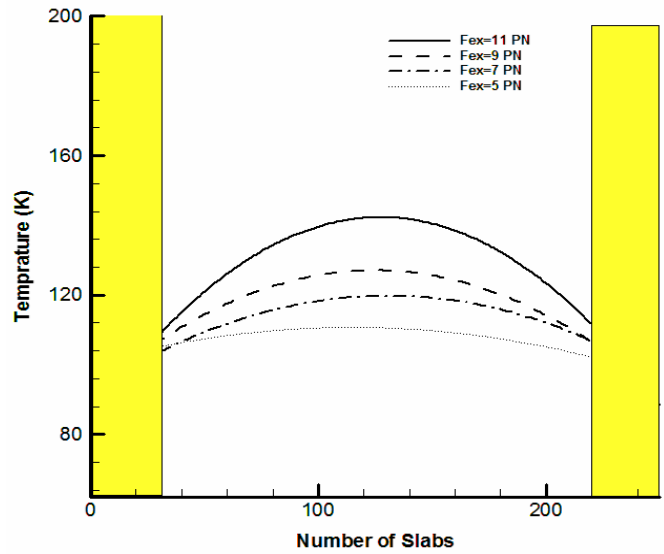

**Figure 7-a:** T=102 K

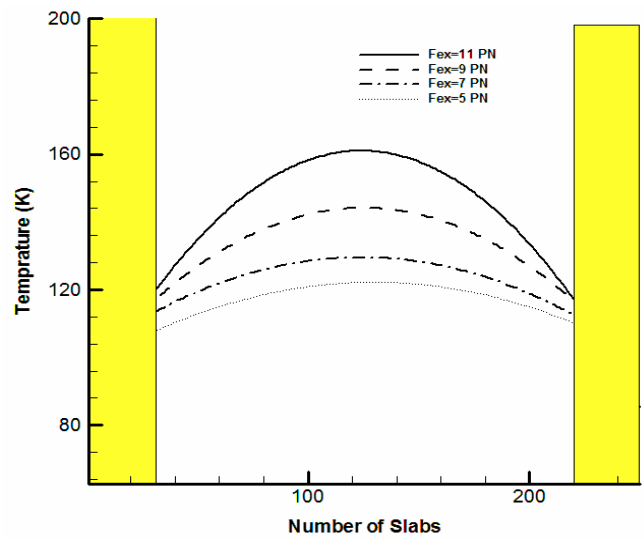
**Figure 7-b:** T=108 K
**Figure 7** Temperature variations for various external forces at different temperature